\documentclass[aps,preprint]{revtex4-1}
\usepackage{eurosym}
\usepackage{bm}
\usepackage{amsfonts}
\usepackage{amsmath}
\usepackage{amssymb}
\usepackage{graphicx}

\begin{document}

\title{On the PBF neutrino losses in superfluid cores of neutron stars}
\author{Lev B. Leinson}
\email{leinson@yandex.ru}

\begin{abstract}
Axial anomalous contributions into neutrino PBF losses due to triplet
pairing of neutrons are still ignored in modeling the evolution of neutron
stars. In this paper, the influence of the anomalous axial contributions
onto the rate of neutron stars cooling is estimated.
\end{abstract}

\keywords{Neutron star, Neutrino radiation, Superfluidity}
\pacs{26.60.-c, 74.20.Fg, 26.30.Jk}
\maketitle

\affiliation{Pushkov Institute of Terrestrial Magnetism, Ionosphere and 
Radiowave Propagation of the Russian Academy of Science (IZMIRAN),\\
108840 Troitsk, Moscow, Russia}


\section{Introduction}

\label{sec:int}
The minimal cooling paradigm \cite{Page2009} suggests that, below the
critical temperature for a triplet pairing of neutrons, the dominant
neutrino energy losses occur from the superfluid neutron liquid in the inner
core of a neutron star. It is commonly believed \cite%
{Tamagaki1970,Hoffberg1970,Baldo1992,Elgaroy1996} that, in this case, the $%
^{3}$P$_{2}$ pairing (with a small admixture of $^{3}$F$_{2}$ state) takes
place with a preferred magnetic quantum number $m_{j}=0$. As derived in \cite%
{Leinson2010}, the neutrino emissivity in this case equals to 
\begin{equation}
Q=\frac{2C_{A}^{2}}{15\pi ^{5}\hbar ^{10}c^{6}}\mathcal{N}_{\nu
}G_{F}^{2}p_{F}M_{n}^{\ast }\left( k_{B}T\right) ^{7}F_{t}\left( \Delta _{%
\mathbf{n}}/T\right) ~,  \label{Q}
\end{equation}%
where the function $F_{t}$ is given by%
\begin{equation}
F_{t}=\int \frac{d\mathbf{n}}{4\pi }y^{2}\int_{0}^{\infty }dx\frac{z^{4}}{%
\left( 1+\exp z\right) ^{2}}.  \label{F}
\end{equation}%
Here the notation is used$\ z=\sqrt{x^{2}+y^{2}}$ with $y=\Delta _{\mathbf{n}%
}/T$,~where the anisotropic energy gap $\Delta _{\mathbf{n}}$ is given by 
\begin{equation}
\Delta _{\mathbf{n}}=\Delta _{0}\left( T\right) \sqrt{1+3\cos ^{2}\theta },
\label{b2}
\end{equation}%
The unit vector $\mathbf{n=p}/p$ defines the polar angles $\left( \theta
,\varphi \right) $ on the Fermi surface.

A comparison of this formula with the expression that was originally obtained
in neglecting the anomalous interactions \cite{Yakovlev1999} allows one to
see that the anomalous contributions completely suppress the vector
channel and also suppress four times the energy losses
through the axial channel.

\section{Neutron star cooling simulation}

\label{sec:model}

For simulations of the thermal evolution of a spherically symmetric NS I
used the NSCOOL code \cite{url} I use the
same NS model which is described in \cite{Page2009} but with a change of
reaction constant $a_{nt}$ in Eq. (11) of this work. 

\begin{figure}[tbp]
\resizebox{0.95\hsize}{!}{\includegraphics{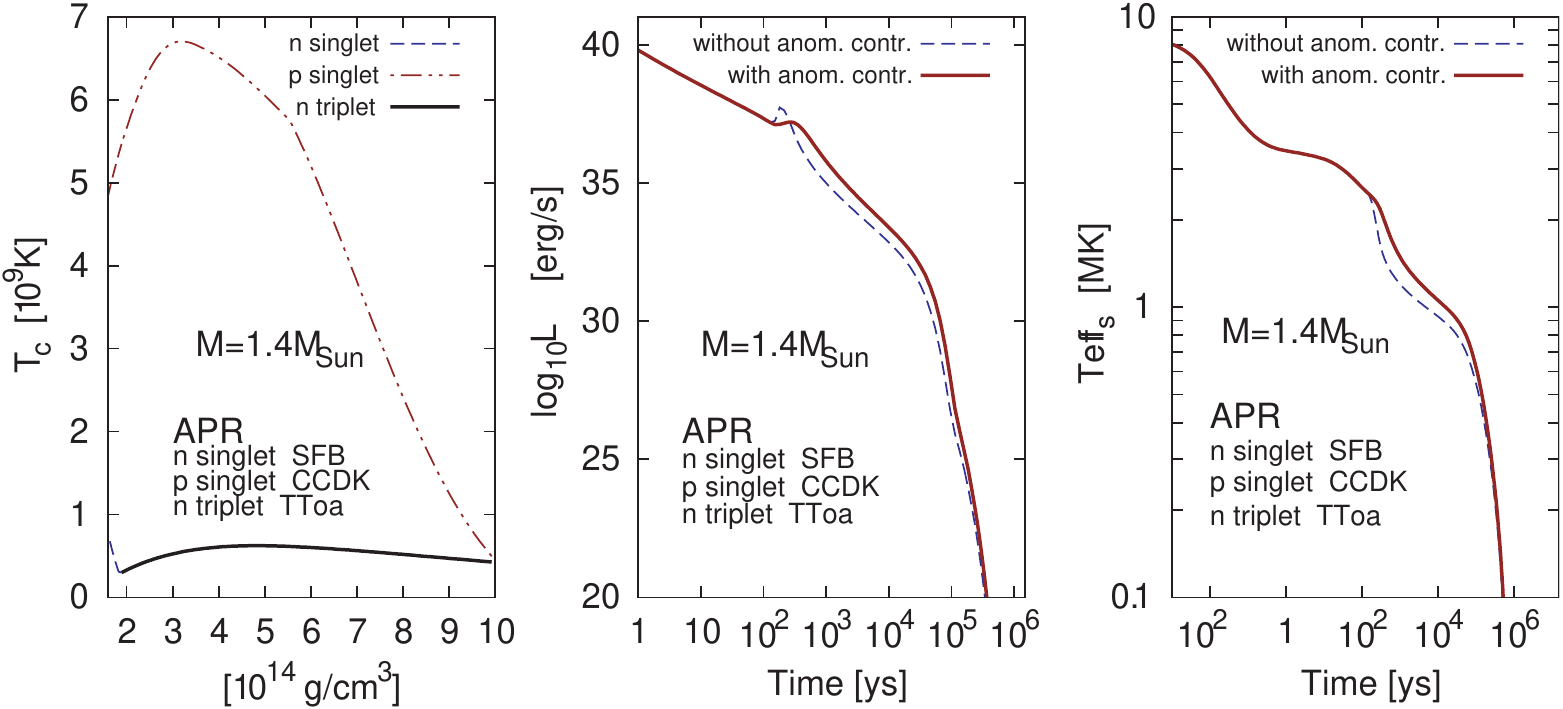}} 
\caption{ (Color online) \textit{Left panel:} Critical temperature $T_{c}$
for neutron superfluidity and proton superconductivity as a function of
matter density for the APR EOS. \textit{Middle panel:} The rate of neutrino
energy losses from NS. \textit{Right panel: }Non-redshifted surface
temperature $T_{s}$ as a function of the NS age. The NS mass is $%
M=1.4\,M_{\odot }$ (iron envelope). The lower cooling trajectory was
obtained for the case when the anomalous weak interactions are discarded, as
it is the case in traditional approach, the upper trajectory was calculated
with inclusion of the anomalous terms.}
\label{fig:trpa}
\end{figure}

\section{Conclusion}

\label{sec:conclusion}

It is necessary to note that though all the calculations are made in a frame
of the used model and the parameters of the mode are known only in order of
magnitude, for example, the critical temperatures for the superfluidity
onset. Nevertheless, a more accurate description of the PBF processes can be
helpful in a treatment of observations. 

\end{document}